\documentclass[review]{elsarticle}
\usepackage{lineno,hyperref}
\modulolinenumbers[5]
\journal{Journal of \LaTeX\ Templates}

\begin{document}
\begin{frontmatter}

\title{ Mott-Hubbard phase transition, gapped electron liquid in insulator state}
\tnotetext[mytitlenote]{M Mott-Hubbard phase transition, gapped electron liquid in insulator state}
\author{Igor N.Karnaukhov}
\address{G.V. Kurdyumov Institute for Metal Physics, 36 Vernadsky Boulevard, 03142 Kiev, Ukraine}
\fntext[myfootnote]{karnaui@yahoo.com}



\begin{abstract}
Within the framework of a mean-field approach the Mott-Hubbard phase transition is considered in the Hubbard and Falicov-Kimball models for half-filled occupation. It is shown that a static $Z_2$-field forms an insulator state on the lattice with a double cell, its strength is determined by the Hubbard interaction. An uniform configuration of the $Z_2$-field corresponds to a gapless spin liquid state, the configuration, at which the lattice with a double cell is formed, corresponds to a gapped fermion liquid, fermions move in this field. Due to the presence of a static field  in an insulator state, a formation of a gapped electron liquid is similar to the gapless Majorana spin liquid in Kitaev's model \cite{AK}. A gap in the spectrum is calculated depending on the magnitude of the Hubbard interaction for the chain, square and cubic lattices. The proposed approach allows us to describe  the Mott-Hubbard phase transition and the insulator state within  the same formalism for an arbitrary dimension for different models.
\end{abstract}

\begin{keyword}
\texttt Hubbard model  \sep Falicov-Kimball model\sep Mott transition
\end{keyword}
\end{frontmatter}

\section{Introduction}

We study the metal-insulator Mott transition within the Hubbard model, using the contact density-density interaction between fermions, so we are talking about the Mott-Hubbard phase transition. In the Mott insulators, the behavior of  an electron liquid is understood as a cooperative many-electron phenomenon, nature of which, unfortunately, is still not clear. The greatest progress has been made in obtaining an exact  solution of the Hubbard chain \cite{LW}, which, however, did not make it possible to understand the peculiarities of the Mott-Hubbard phase transition for an arbitrary dimension. This simple model (at first glance) exhibits a rich phase diagram with a striking set of phases and regimes.

We still do not know a criterion of the Mott-Hubbard phase transition.
So, realization of perfectly nested Fermi surfaces at a half-filled occupation \cite{C,IK1} can not be criterion of the Mott-Hubbard phase transition.
Firstly, the interaction is not taken into account in this case, and in the limit of strong interaction it is this interaction that determines a dielectric gap, secondly, if perfect nesting is not realized, this does not mean that the insulator state cannot be realized either. The criterion must take into account the parameters of the model such, a constant of the interaction, filling, hopping integral for the next-nearest neighbors (for an extended model).
Even at a half-filled occupation, the calculations according to the Hubbard model give different results, and significantly  (see review \cite{1}). The reason is simple, we still  have not understood the peculiarities of behavior of the Mott-Hubbard transition.

We declare that the Mott-Hubbard phase transition can be described as formation a insulator state on the lattice with a double cell, a electron liquid in insulator state is similar to the  Majorana spin liquid. We have considered a two-component mean field, one of which corresponds to a static $Z_2$-field, its amplitude is determined by an interaction strength. Turns out that the energy minimum is achieved by an uniform $Z_2$-field configuration, which corresponds to a lattice with a double cell. An electron liquid is free electrons moved in a static $Z_2$-field. In contrast to the Majorana spin liquid \cite{AK} an electron liquid in an insulator state is gapped and non-topological. An uniform configuration of a $Z_2$-field forms the insulator state with a spin- or charge-antiferro-order. A stability of the configuration of the $Z_2$-field is a criterion of the realization of an insulator state at the Mott-Habbard phase transition.

\section{The Hubbard model}
The Hamiltonian of the Hubbard model has the well-known form ${\cal H}_{Hub} ={\cal H}_{0} +{\cal H}_{int} $
\begin{eqnarray}
&&{\cal H}_{0} = - \sum_{<i,j>} \sum_{\sigma =\uparrow,\downarrow} c^\dagger_{i,\sigma} c_{j,\sigma}-t \sum_{<<i,j>>} \sum_{\sigma =\uparrow,\downarrow} c^\dagger_{i,\sigma} c_{j,\sigma}
,\nonumber\\&&
{\cal H}_{int}={U} \sum_{j}n_{j,\uparrow}n_{j,\downarrow},
\label{eq-Hub}
\end{eqnarray}
where $c^\dagger_{j,\sigma}$ and $c_{j,\sigma}$ are the Fermi operators that determine the electrons on a site \emph{j} ($\sigma=\uparrow,\downarrow$ is the spin of the electron),  $n_{j,\sigma} =c^\dagger_{j,\sigma}c_{j,\sigma}$ denote the density operators, 
${U}$ is a magnitude of an on-site Coulomb repulsion, 1, $t$  and $z_1$, $z_2$ are the hopping integrals of electrons  between nearest, next-nearest neighbors and the numbers of nearest, next to nearest  neighbors, respectively. Below we consider the models for a simple symmetry of the dimension $D$: the chain, square and cubic lattices.

\subsection{A repulsion interaction}

In section "Methods" we have shown that $\Lambda-$ and $\lambda$-components of the mean field are determined by the magnitude of the on-site Hubbard interaction. A static $\emph{Z}_2$-field, connected with $\Lambda$-component, forms the ground state in an insulator state. Numerical analysis of Eqs (7) shows that, at half-filled occupation for $t=0$, arbitrary  $\Lambda$, $\lambda$,  an uniform configuration of the $\emph{Z}_2$-field corresponds to the ground state of   an electron liquid in an insulator state \cite{AK,Lieb,KS,K0}. In contrast to the  gapless Majorana spin liquid \cite{AK}, where in the ground state a free $Z_2$-field configuration with the same quantum flux through the hexagon is realized and a lattice cell does not change, this configuration corresponds to lattice with a double cell. The electron liquid state with a double cell in gapped.
At half-filled occupation for $t=0$ an insulator state  is stable one for arbitrary values of $\Lambda$ and $\lambda$.
With increasing doping  the double lattice cell  becomes unstable, a small doping does not kill this $Z_2$-field configuration.

Below we will analyse the phase transition at half-filling  in detail.
At T=0K the system of equations which correspond to a saddle point of  the action $S$ (8) has the following form
\begin{eqnarray}
\frac{\Lambda}{U}- \frac{\Lambda}{4 N}\sum_\textbf{k} \left(   \frac{1}{\varepsilon_+(\textbf{k})}+  \frac{1}{\varepsilon_-(\textbf{k})}       \right) =0,\nonumber \\
\frac{\lambda}{U}- \frac{1}{4 N}\sum_\textbf{k} \left(   \frac{\lambda +|w(\textbf{k})|}{\varepsilon_+(\textbf{k})}+  \frac{\lambda -|w(\textbf{k})|}{\varepsilon_-(\textbf{k})} \right) =0,
\label{eq-Eqs}
\end{eqnarray}
where $N$ is the number of lattice sites, the components of the $\lambda$-field are determined by the value of $U$. We also taking into account that the lattice in insulator state has a double cell.

\begin{figure}[tp]
     \centering{\leavevmode}
\begin{minipage}[h]{.35\linewidth}
\center{
\includegraphics[width=\linewidth]{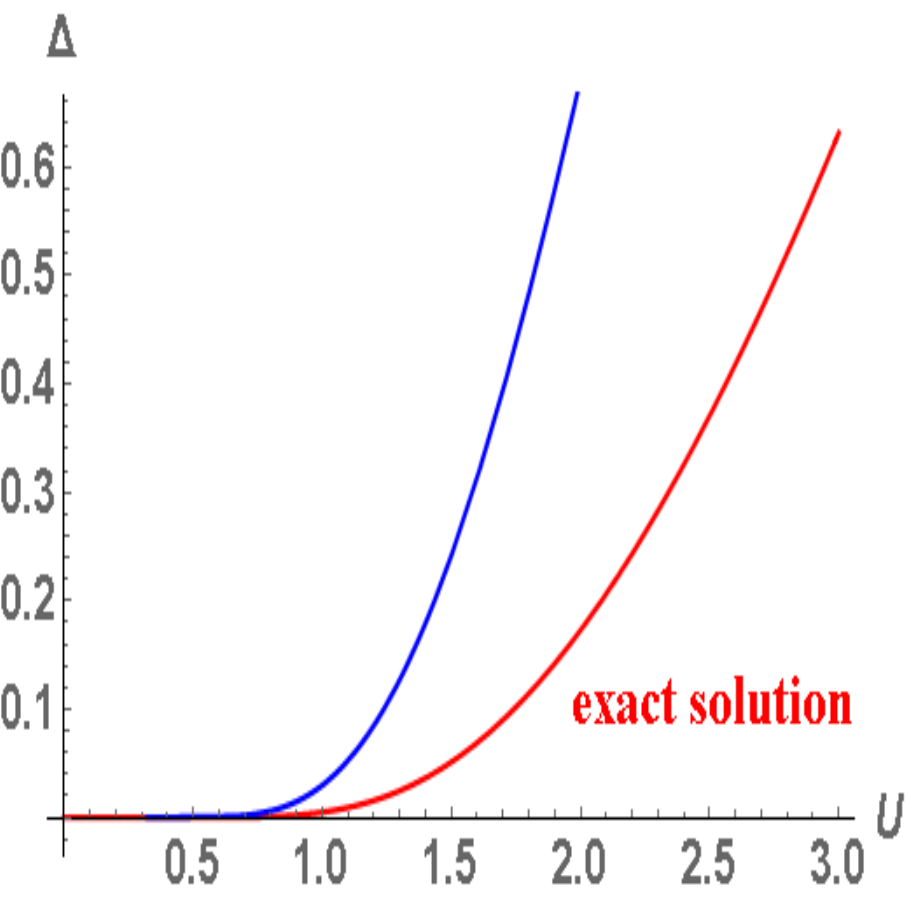} a)\\
                 }
   \end{minipage}
\begin{minipage}[h]{.34\linewidth}
\center{
\includegraphics[width=\linewidth]{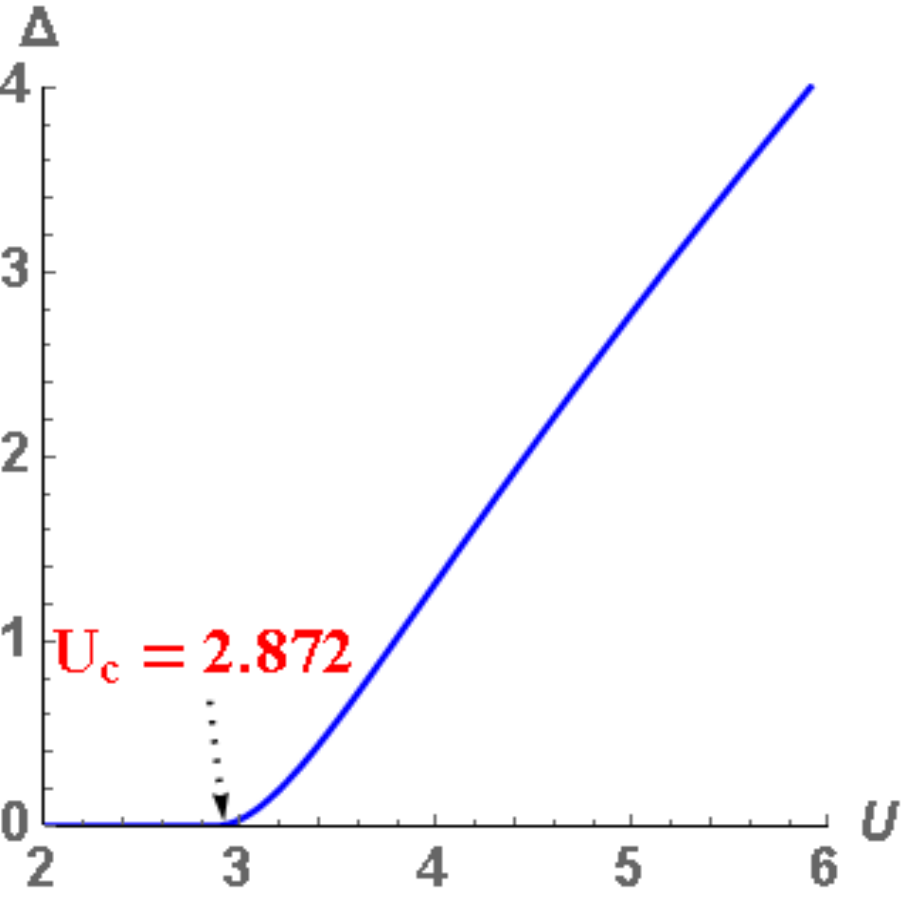} b)\\
                  }
   \end{minipage}
   \begin{minipage}[h]{.28\linewidth}
\center{
\includegraphics[width=\linewidth]{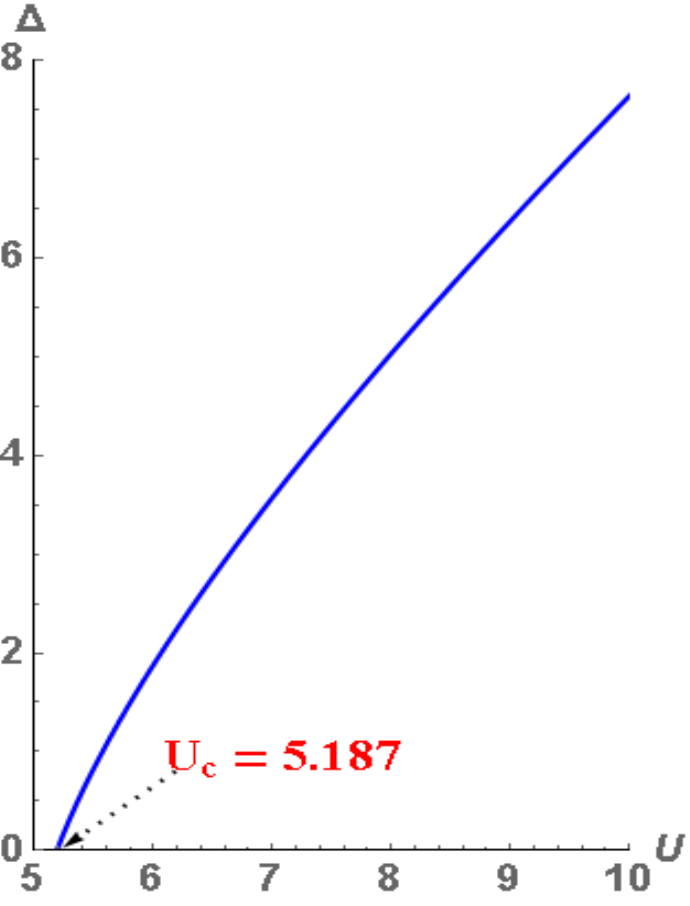} c)\\
                  }
   \end{minipage}
\caption{(Color online)
The gap $\Delta$ as a function of $U$  calculated in the Hubbard model at half-filled occupation for the chain a) (the exact solution is presented for comparison), square b), $U_c=2.872$ and cubic c)  $U_c=5.187$ lattices.
 }
\label{fig:1}
\end{figure}

Numerical analysis of Eqs (2) shows that, the solutions for $\Lambda$ and $\lambda$ are $\Lambda\neq0$, $\lambda=0$ and $\Lambda=0$, $\lambda\neq 0$,
the first solution  $\Lambda\neq0$, $\lambda=0$ corresponds to the minimum of the energy. This solution conserves the particle numbers of electrons with different spin, that takes place in the Hamiltonian (1). A solution $\Lambda\neq 0$ and $\lambda\neq 0$ is not realized for given $U$.
$\Lambda_\textbf{j}$ variable is associated with the  operator $n_{\textbf{j},\uparrow}-n_{\textbf{j},\downarrow}$, thus the spin  antiferro order is realized in the insulator state.
We will analyze the formation of a gap in the electron spectrum at half-filled occupation and focus our attention on opening the gap with increasing the interaction strength in the chain and  square, cubic lattices.
In the case of the weak interaction ${U}\to 0$, we obtain the solution for the gap $\Delta\simeq G \exp(-2\pi/{U})$ for the chain \cite{IK1}, where $G$ is the pre-exponential factor determined by cutoff  and  $\Delta \to U$ in strong coupling limit.
We can compare with an exact solution of the Hubbard chain, where  $\Delta = \frac{16}{{U}} \int_1^\infty d{x} \frac{\sqrt{{x}^2-1}}{\sinh(2\pi{x} /{U})}$ \cite{LW,AA}, and  the asymptotic expressions at ${U}\to 0$  $\Delta=8\sqrt{{U}}/\pi \exp(-2\pi/{U})$ (see in Fig \ref{fig:1}a)) and for large $U$ $\Delta = U-4+8\ln{2}/U$.
To visually illustrate the calculations of the dependence of the gap value in the quasi-particle spectrum depending on the interaction strength.
In Figs 1 the calculations of gap  are presented for the chain a), square b) and cubic c)  lattices. In Fig 1 a) the exact solution is shown also  for comparison. A minimal value of an on-site repulsion $U_c$ at which the gap in the spectrum opens is equal to 2.872 for square and 5.187 for cubic lattices. The value of the interaction strength $U_c$  is an important characteristic of the phase transition.

\subsection{Nearest- and next-nearest hoppings}

With increasing the value of the hopping integral of electrons between next-nearest neighbors $t$, the lattice with a double cell becomes unstable. Thus, at half-filled occupation in the region of low $\Lambda$-values, the state becomes unstable at $\frac{z_2t}{z_1 }>\frac{1}{2}$, at the same time it is stable for arbitrary $\Lambda$ and $\lambda$ at $\frac{z_2t}{z_1 }<\frac{1}{2}$. For the simple structures, such chain, square and cubic lattices $z_1=z_2$ the criterion of the stability of an insulator state with a double lattice cell is reduced to $t<\frac{1}{2}$.
The behavior of an electron liquid in an insulator state is different at small $t<t_c$ and large $t>t_c$. The value $t_c$ is determined as maximal value of the hopping integral for electrons between next-nearest neighbor lattice sites at which the gap in the quasi-particle spectrum opens for $\Lambda \to 0$, so for the chain $t_c=0.32$ and $t_c=0.125$ for the square and cubic lattices. At $t>t_c$ the gap opens for a finite value of $\Lambda$.

The quasi-particle spectrum is not symmetric about zero energy, the energies $\pm\varepsilon_{\pm}(\textbf{k})$ (9) are shifted by the term $\delta\varepsilon (\textbf{k})=-2t\sum^D \cos(2k_\alpha)$, namely  $\delta\varepsilon (\textbf{k})\pm\varepsilon_{\pm}(\textbf{k})$. A term $\delta \varepsilon (\textbf{k})$ enters additively into the formula for the excitation energies, so Eqs (2) do not change when taking into account $t$. And as a result, the value of the gap $\Delta$
does not depend on $t$ for $t<t_c$, wherein the chemical potential is shifted by a value $2Dt$ at $\Lambda \to 0$.
The value $U_c$ as a function of $t$, where $t$ changes in an interval $0<t<1/2$, is shown in Fig.2. As expected, the value of $U_c$ increases with $t$, but only for $t>t_c$. The behavior of the electron liquid in the chain also differs sharply from the 2D and 3D model, an insulator state is also realized  for arbitrary $t<\frac{1}{2}$.

\begin{figure}[tp]
     \centering{\leavevmode}
\begin{minipage}[h]{.35\linewidth}
\center{
\includegraphics[width=\linewidth]{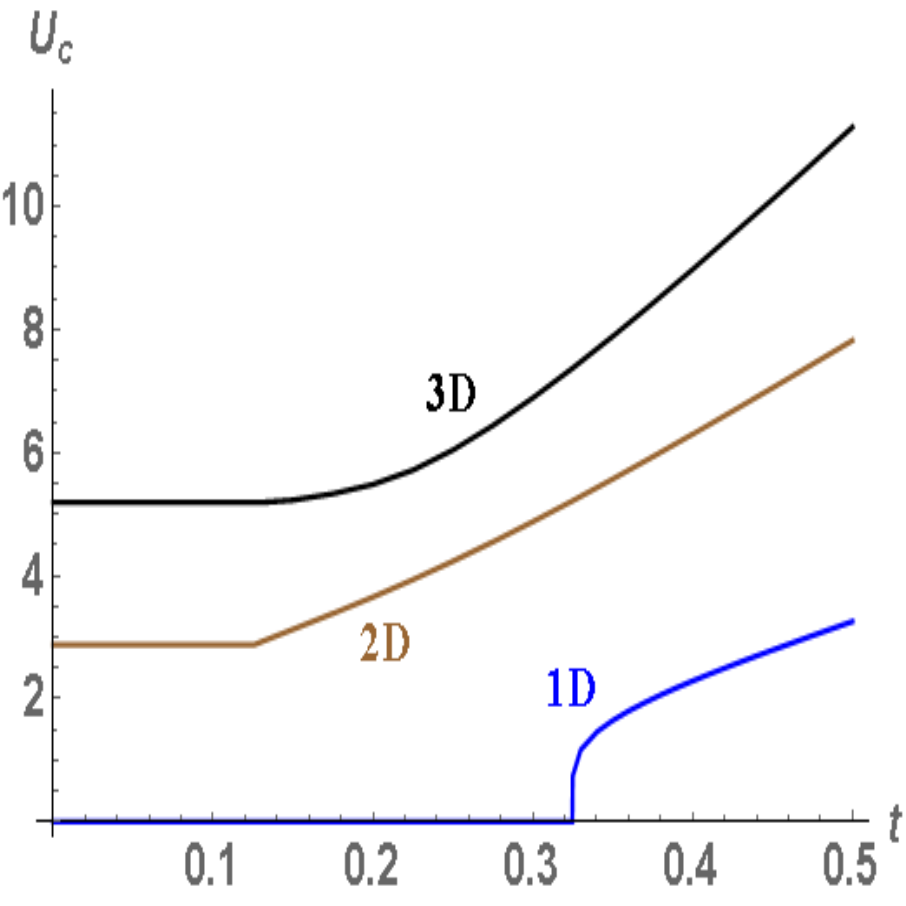}\\
                 }
   \end{minipage}
\caption{(Color online)
Critical value $U_c$  as a function of $t$  calculated in the Hubbard model with nearest- and next nearest hoppings at a half-filled occupation for the chain (1D), square (2D) and cubic (3D) lattices.
 }
\label{fig:2}
\end{figure}

\subsection{An attractive interaction}

The systems of Eqs (7) and (10) have the same structure, so the results of numerical analysis show, that the $Z_2$-field with variables $\Omega_\textbf{j} = \pm \Omega$ at $\omega_\textbf{j} = \omega$  and $t =0$ corresponds to the minimum of the energy  for an uniform configuration of $\Omega_{\textbf{j}}=\Omega $, $\Omega_{\textbf{j+a}}=-\Omega$. At $t=0$ this configuration corresponds to the minimum of the energy for arbitrary values of $\Omega$ and $\omega$.
$\Omega\neq 0$ defines the $Z_2$-field, which  forms the phase state of an electron liquid on the lattice with a double cell, the spectrum of the quasi-particle excitations is degenerated with energies $ E_{\pm}(\textbf{k})=\pm \sqrt{4 \Omega^2 + |\omega|^2+ |w(\textbf{k})|^2}$. $\Omega-$  and $\omega-$components of the mean field determine the gap in the spectrum, their solutions  are defined by the system of the self-consistent equations.
$\Omega_\textbf{j}$ variable is related to the density operator $n_{\textbf{j},\uparrow}+n_{\textbf{j},\downarrow}$, thus the charge antiferro order is realized at $\Omega \neq 0$.
In contrast to the repulsive Hubbard interaction, where nontrivial solution with  $\Lambda\neq 0$ and $\lambda \neq 0$ does not satisfy self-consistent equations (2), a solution $|\omega | = 2\Omega$ also satisfies the corresponding system of equations.
At $|\omega|=2\Omega$ the system of equations is reduced to the following equation
\begin{eqnarray}
\frac{\omega}{U}- \frac{\omega}{2 N}\sum_\textbf{k}  \frac{1}{\varepsilon (\textbf{k})} =0,
\label{eq-Eqs}
\end{eqnarray}
where $\varepsilon (\textbf{k}) =\sqrt{2 |\omega|^2 +|w(\textbf{k})|^2}$.

The energy of an electron liquid is  a monotonically decreasing function of $\omega$, value of $U$ is a monotonically increasing function of $\omega$ for arbitrary model dimension.
A state, in which the charge and superconducting orders are mixed,  has a higher energy (for a fixed $U$, it corresponds to an energy with $\omega$ to $\sqrt{2}$ less), than a state with only charge or superconducting order, so it is not stable. The energies of states with the  charge or superconducting order are degenerate, nevertheless, the charge order is most likely to be realized, since the $Z_2$-field in this case does not break the symmetry of the Hamiltonian (1), reflects global charge conservation. In this case the effective Hamiltonian is closer to the initial one (1), so it more adequately describe the ground state and low-energy excitations. These arguments do not work for non-equilibrium state of the Hubbard model at half-filling \cite{KSU}.
Should be noted also that the value of $U_c$ at which the gap opens is the same as for the model with the repulsive interaction, therefore we can not say about the superconducting state in the case of a weak interaction.

\section{The Falicov-Kimball model}

Recall  the Hamiltonian of the Falicov-Kimball model ${\cal H}_{FK}$
\begin{eqnarray}
&&{\cal H}_{FK} = -\sum_{<i,j>} c^\dagger_{i} c_{j}+\epsilon  \sum_j d^\dagger_j d_j+U \sum_{j}c^\dagger_j c_j d^\dagger_j d_j,
\label{eq-FK}
\end{eqnarray}
where $c_{j}$ and $d_{j}$ are the spinless Fermi operators that determine the s- and d-fermion states on a site \emph{j}, $\epsilon$ is the energy of one-particle d-states, $U>0$ defines  a contact interaction between s- and d-fermion states.

At half-filled occupation when the energy of one-particle d-states lies on the Fermi energy or into a gap of the quasi-particle spectrum, the gap opens and an insulator state is realized \cite{DG1,DG2}. The two component $\lambda$-field is determined by an analogous for the Hubbard model with repulsion interaction, where $c_{j,\uparrow} \to c_{j}$, $c_{j,\downarrow} \to d_j$ with hopping integrals of fermions between the nearest-neighbor sites  equaled to one for s- and zero for d-fermions. In an insulator state a $\Lambda$-component of the mean field realises an electron liquid state on the lattice (or chain in one dimension) with a double cell. The spectrum of the quasi-particle excitations is symmetric about zero energy at $\epsilon=0$ and includes four branches
\begin{equation}
\epsilon^2_{\pm}(\textbf{k})=4 \Lambda^2 + |\lambda|^2+\frac{1}{2}|w(\textbf{k})|^2 \pm \frac{1}{2} |w(\textbf{k})|\sqrt{4|\lambda|^2+|w(\textbf{k})|^2}.
\label{A3}
\end{equation}

\begin{figure}[tp]
     \centering{\leavevmode}
\begin{minipage}[h]{.3\linewidth}
\center{
\includegraphics[width=\linewidth]{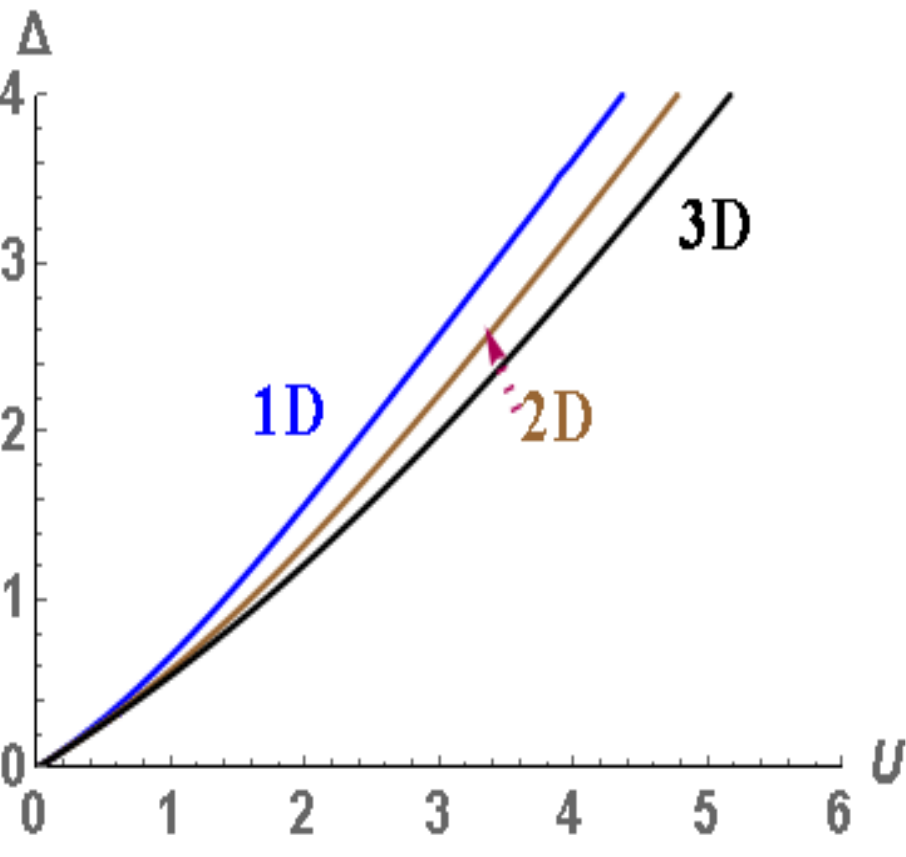} a) \\
                 }
   \end{minipage}
   \begin{minipage}[h]{.3\linewidth}
\center{
\includegraphics[width=\linewidth]{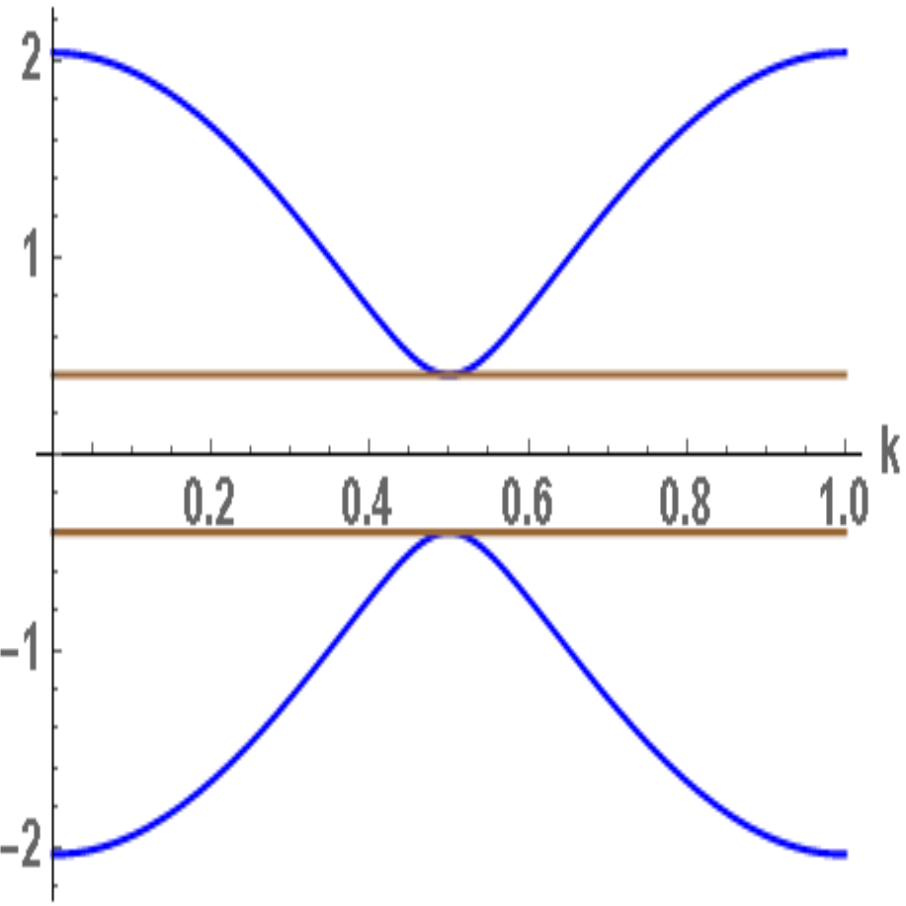}  b) \\
                 }
   \end{minipage}
\caption{(Color online)
The gap $\Delta$ as a function of $U$  calculated in the Falicov-Kimball  model at half-filing occupation and $\epsilon=0$ for the chain 1D, square 2D and cubic 3D  lattices a). The spectrum of excitations calculated for the chain at $\epsilon=0$, $\Lambda =0.2$, $U=1.19$ b).
 }
\label{fig:3}
\end{figure}

In a saddle point approximation the amplitudes of the mean field are determined by equations which are similar to (2).
The interaction strength $U$ determines a $\Lambda$-value, the value of which does not depend on  $\epsilon$  at half-filled occupation and is calculated from the first equation in (2) with the change of $\varepsilon_{\pm}(\textbf{k}) \to \epsilon_{\pm}(\textbf{k})$.
Numerical analysis shows that
the solution with $\Lambda \neq 0$ and $\lambda\neq0$ is not realized (as in the Hubbard model with repulsion),
the solution $\Lambda\neq 0 $, $\lambda =0$ corresponds to the minimum of the energy, and the spectrum (5) includes dispersion $\epsilon_s(\textbf{k})= \pm \sqrt{4\Lambda^2+|w(\textbf{k})|^2}$ and dispersionless $\epsilon_g=\epsilon \pm 2\Lambda$ branches of excitations. When the energy of one-particle d-states lies into the gap, this level splits.
Results of calculations of the gap $\Delta =4\Lambda$ at $\epsilon=0$ are shown in Fig \ref{fig:3}a) for chain, square and cubic lattices.
The spectrum of quasi-particle excitations calculated for the chain at $\epsilon =0$, $\Lambda =0.2$, $U=1.19$ is shown in Fig \ref{fig:3}b) for illustration. This solution conserves the total numbers of s- and d-fermions (separately) in an insulator state, as only $\lambda$-component of the field leads to hybridization between s- and d-states and breaks these particle number conservations. The spectrum of d-states remains dispersionless, it splits due to an on-site interaction. The value of gap $\Delta$ decreases with increasing $|\epsilon|$ $\Delta=4\Lambda-|\epsilon|$ and $\Delta \to 0$ at $|\epsilon| \to 4\Lambda$ for half-filled occupation. A stability of an uniform configuration of  the $\emph{Z}_2$-field, which corresponds to a double cell, depends on  filling and value of $\epsilon$.
So at half-filling and $|\epsilon |\gg 1$ an free configuration with $\Lambda_\textbf{j}=\Lambda$ corresponds to minimum of energy. From numerical analysis it follows that at $|\epsilon|\leq 2\Lambda$  an uniform configuration which leads to the lattice with a double cell is stable, with increasing $|\epsilon|$, the phase state becomes unstable at $|\epsilon|> 2\Lambda$. Unfortunately, we cannot say for what value of $2\Lambda<|\epsilon|<4\Lambda$ another stable configuration of the $\emph{Z}_2$-field is realized and what is this field configuration.

\section{Conclusions}
In the Hubbard and Falicov-Kimball models the Mott-Hubbard phase transition is realized according to the scenario of the Majorana spin liquid, first considered by Kitaev \cite{AK}. In contrast to the Kitaev model this fermion liquid is non-topological gapped one.  At half-filled occupation an uniform configuration of this field, which corresponds to a lattice with a double cell, responds to minimum of energy.
Fermions move in a static $\emph{Z}_2$-field, which is formed due to the on-site interaction between particles. Nontrivial solution for the amplitude of the mean field, which defines $\emph{Z}_2$-field, conserves the particle numbers, forms the gapped spectrum of quasi-particle excitations, leads to an insulator state. The results of calculations show that the concept of the state of the Majorana spin liquid is productive for describing the insulator state at the Mott phase transition.

\section{Acknowledgments}

It is the third month of the war between Ukraine and Russia, the author thanks the Ukrainian army and the Ukrainian people, who courageously
defend Ukraine. The author also thanks those russian scientists who understand the horrors of this war for Ukraine and support ukrainian
colleagues.

\section{Methods\\
Mean field approach with a two component $\lambda$-field
}
\subsection{The Hubbard model with a repulsion interaction}

Using the following relations between the operators we redefine the interaction term in the model Hamiltonian (1) as
$n_{\textbf{j},\uparrow}n_{\textbf{j},\downarrow}+n_{\textbf{j},\downarrow}n_{\textbf{j},\uparrow}=-(n_{\textbf{j},\uparrow}-n_{\textbf{j},\downarrow})^2+ n_{\textbf{j},\uparrow}+n_{\textbf{j},\downarrow}$ and
$n_{\textbf{j},\uparrow}n_{\textbf{j},\downarrow}=-\chi^\dagger_{\textbf{j}}\chi_{\textbf{j}}$, where $\chi_\textbf{j}= c^\dagger_{\textbf{j},\uparrow}c_{\textbf{j},\downarrow}$ and $n^2_{\textbf{j},\sigma}=n_{\textbf{j},\sigma}$.
Introducing a two parametric $\lambda$-field we can redefine the interaction term due to the Hubbard-Stratonovich transformation
$2\frac{\Lambda^2_\textbf{j}}{U}+2\Lambda_\textbf{j}(n_{\textbf{j},\uparrow}-n_{\textbf{j},\downarrow})+
\frac{|\lambda_{\textbf{j}}|^2}{{U}}+\lambda_{\textbf{j}}\chi^\dagger_{\textbf{j}}+
\lambda^\ast_{\textbf{j}}\chi_{\textbf{j}}$.  We define the interaction term in the action  $S_{int}$  introducing the action ${S}_{0}$, which is determined by the Hamiltonian of the non-interacting fermions
\begin{equation}
{ S}_{int}={S}_{0}+ 2\sum_{\textbf{j}}\left(\frac{\Lambda^2_{\textbf{j}}}{U}+\Lambda_{\textbf{j}}(n_{\textbf{j},\uparrow}-n_{\textbf{j},\downarrow})\right)+
\sum_{\textbf{j}}\left(\frac{|\lambda_{\textbf{j}}|^2}{{U}}+\lambda_{\textbf{j}}\chi^\dagger_{\textbf{j}}+ \lambda^\ast_{\textbf{j}}\chi_{\textbf{j}}\right).
 \label{A1}
\end{equation}

The canonical functional is defined as ${\cal L}=\int {\cal D}[\Lambda,\lambda] \int {\cal D}[n_\uparrow,n_\downarrow,\chi\dagger,\chi] e^{-S}$, where the action $S=\frac{2}{U}\sum_\textbf{j} \Lambda^2_\textbf{j} + \frac{1}{U}\sum_{\textbf{j}}|\lambda_{\textbf{j}}|^2
+ \int_0^\beta d\tau \Psi^\dagger (\tau)[\partial_\tau  + {\cal H}_{eff}]\Psi (\tau)$ with
$ {\cal H}_{eff}=  {\cal H}_0 +2\sum_\textbf{j} \Lambda_\textbf{j}(n_{\textbf{j},\uparrow}-n_{\textbf{j},\downarrow})+ \sum_{\textbf{j}}(\lambda_{\textbf{j}}\chi^\dagger_{\textbf{j}}+\lambda^\ast_{\textbf{j}}\chi_{\textbf{j}})$, and $\Psi(\tau)$ is the wave function. We expect $ \Lambda_\textbf{j}$ and  $ \lambda_ {\textbf{j}} $ to be independent of $ \tau $ due to translational invariance.
Two component $\lambda$-field leads to hybridization between electron states with different spin (due to the $\lambda$-amplitude) and to formation of a static $\emph{Z}_2$-field (due to the $\Lambda$-amplitude).

For investigation of the ground state of the system, determined by the effective Hamiltonian ${\cal H}_{eff}$, let us consider the equations for the amplitudes of the wave function $\psi_\sigma (\textbf{j})c^\dagger_{\textbf{j},\sigma}$ and $\psi_{-\sigma}(\textbf{j}) c^\dagger_{\textbf{j},-\sigma}$  with the energy $\varepsilon$ in the $\emph{Z}_2$-field:
\begin{eqnarray}
&&(\varepsilon -2\Lambda_\textbf{j}) \psi_{\sigma}(\textbf{j})-\sum_{\textbf{a}}\psi_{\sigma}(\textbf{j+a})-
t\sum_{\textbf{b}}\psi_{\sigma}(\textbf{j+b}) + \lambda^*_\textbf{j}\psi_{-\sigma}(\textbf{j})=0, \nonumber\\
&&\overline{}
(\varepsilon +2\Lambda_\textbf{j}) \psi_{-\sigma}(\textbf{j})-\sum_{\textbf{a}}\psi_{-\sigma}(\textbf{j+a}) - t\sum_{\textbf{b}}\psi_{-\sigma}(\textbf{j+b}) +\lambda_\textbf{j}\psi_{\sigma}(\textbf{j})=0,
\end{eqnarray}
where the summations over the nearest-neighbor lattice sites with the coordinates $\textbf{j+a}$ and  the next nearest-neighbor lattice sites with the coordinates $\textbf{j+b}$. The action is a quadratic function of $\Lambda_j$,  we consider the two values of the $\Lambda$-field $\Lambda_j=\pm \Lambda$ chaotically located at lattice sites, for $\lambda_\textbf{j}=\lambda$.
As we noted above, the variables  $\Lambda_\textbf{j}=\pm \Lambda$  are identified with a static $Z_2$-field determined on the lattice sites, the band electrons move in this static field. Detailed numerical analysis of Eqs (7) shows, that for $t=0$ an uniform configuration with $\Lambda_\textbf{j} =\Lambda$ and $\Lambda_\textbf{j+a}=-\Lambda$ for all variables corresponds to the ground state of an electron liquid described by Eqs (7) \cite{AK,Lieb,KS}. The action $S$ and ${\cal H}_{eff}$ can  be solved for an uniform configuration of a static $Z_2$-field, one can integrate out the fermionic contribution to obtain the action $S$ per atom, where $\omega_n ={T}(2n+1)\pi$ are the Matsubara frequencies
\begin{equation}
\frac{S}{\beta}=-\frac{{T}}{N}\sum_{\textbf{k}}\sum_n \ln [(\omega^2_n+\varepsilon^2_+(\textbf{k}))( \omega^2_n+\varepsilon^2_-(\textbf{k}))]+2\frac{\Lambda^2}{{U}}+\frac{|\lambda|^2}{{U}},
 \label{A2}
\end{equation}
where
\begin{equation}
\varepsilon^2_{\pm}(\textbf{k})=4 \Lambda^2 + (|\lambda| \pm |w(\textbf{k})|)^2,
\label{A3}
\end{equation}
where $w(\textbf{k})=\sum^D(1+\exp(i k_\alpha))$, $\textbf{k}=(k_x,k_y,k_z)$ is the wave vector.

We define the solutions for the amplitudes of the $\lambda$-field in a saddle point approximation for the functional $ {\cal L}$,
the minimal action $ S $ will dominate if $\Lambda$ and $\lambda$ satisfy the following equations $\partial S/\partial\Lambda =0$ and $\partial S/\partial\lambda =0$.

\subsection{The Hubbard model with an attractive interaction}

In the Hubbard model with an attractive interaction we will use the following relations for the operators
$-n_{\textbf{j},\uparrow} n_{\textbf{j},\downarrow} - n_{\textbf{j},\downarrow} n_{\textbf{j},\uparrow}=
-(n_{\textbf{j},\uparrow}+n_{\textbf{j},\downarrow})^2+ n_{\textbf{j},\uparrow}+n_{\textbf{j},\downarrow}$ and
$-n_{\textbf{j},\uparrow} n_{\textbf{j},\downarrow}=-\gamma^\dagger_\textbf{j} \gamma_\textbf{j}$ with
$\gamma_\textbf{j} = c_{\textbf{j},\downarrow}c_{\textbf{j},\uparrow}$. The corresponding term to the action has the following form\
$\sum_{\textbf{j}}[2\frac{\Omega^2_\textbf{j}}{|U|}+2\Omega_\textbf{j}(n_{\textbf{j},\uparrow}+n_{\textbf{j},\downarrow})+
\frac{|\omega_{\textbf{j}}|^2}{{|U|}}+\omega_{\textbf{j}}\gamma^\dagger_{\textbf{j}}+
\omega^\ast_{\textbf{j}}\chi_{\textbf{j}}]$.

Similarly to the case of repulsive interaction,  consider the equations for the amplitudes of the wave function $\psi_\sigma (\textbf{j})c^\dagger_{\textbf{j},\sigma}$ and $\varphi_{-\sigma}(\textbf{j}) c_{\textbf{j},-\sigma}$  with the energy $\varepsilon$ in the $\emph{Z}_2$-field, which is defined by a variable $\Omega_\textbf{j}=\pm \Omega$ on each lattice site:
\begin{eqnarray}
&&(\varepsilon +2\Omega_\textbf{j}) \psi_{\sigma}(\textbf{j})- \sum_{\textbf{a}}\psi_{\sigma}(\textbf{j+a}) -t\sum_{\textbf{b}}\psi_{\sigma}(\textbf{j+b}) +\omega^*_\textbf{j}\varphi_{-\sigma}(\textbf{j})=0, \nonumber\\
&&\overline{}
(\varepsilon -2\Omega_\textbf{j}) \varphi_{-\sigma}(\textbf{j})+\sum_{\textbf{a}}\varphi_{-\sigma}(\textbf{j+a})+ t\sum_{\textbf{b}}\varphi_{-\sigma}(\textbf{j+b}) +\omega_\textbf{j}\psi_{\sigma}(\textbf{j})=0,
\end{eqnarray}
From the numerical analysis of the stability  of the electron liquid, described by Eqs (10), already traditional conclusion follows that for $t=0$  and $\omega_j\textbf{}=\omega$ an uniform configuration with $\Omega_\textbf{j} =\Omega$ and $\Omega_\textbf{j+a}= -\Omega $ for all variables $\Omega_\textbf{j}$ corresponds to the ground state. At half-filled occupation, the phase state with a double lattice cell is stable for arbitrary values of $\Omega$ and $\omega$. This makes it possible to study the phase state of the electron liquid in the case of an attractive Hubbard interaction.


\end{document}